\documentstyle[aaspptwo]{article}
\tighten
\slugcomment{to appear in ApJ }

\newcommand{\beq}{\begin{equation}}
\newcommand{\eeq}{\end{equation}}
\newcommand{\beqa}{\begin{eqnarray}}
\newcommand{\eeqa}{\end{eqnarray}}

\newcommand{\ea}{{\it et al. }}

\begin{document}
\title{ ROTATION, DIFFUSION, AND OVERSHOOT IN THE
SUN: EFFECTS ON THE OSCILLATION
FREQUENCIES AND THE NEUTRINO FLUX}

\author{Brian Chaboyer\altaffilmark{1}, P.~Demarque\altaffilmark{2},
D.B.~Guenther\altaffilmark{3} \& M.H.~Pinsonneault\altaffilmark{4} }

\altaffiltext{1}
{Canadian Institute for Theoretical Astrophysics, 60 St. George
St., Toronto, Ontario, Canada M5S 1A7 \hfill \\
E-Mail: chaboyer@cita.utoronto.ca}

\altaffiltext{2}{Center for Solar and Space Research, Department of Astronomy,
Yale University, Box 208101, New Haven,
CT 06520-8101\hfill  \\E-Mail: demarque@astro.yale.edu}

\altaffiltext{3}{Department of Astronomy and Physics, Saint Mary's University,
Halifax, Nova Scotia, Canada, B3H 3C3\hfill \\
E-mail:  guenther@romana.stmarys.ca}

\altaffiltext{4}{Department of Astronomy,Ohio State University,
174 W. 18th Ave. Columbus, OH 43210-1106\hfill \\
E-Mail: pinsono@payne.mps.ohio-state.edu}

\begin{abstract}
We have studied the importance of the combined effects of rotation,
diffusion, and convective overshoot on the p-mode oscillation spectrum
and the neutrino flux of the standard solar model. To isolate the
various physical affects included in the new rotation plus diffusion
models we also constructed solar models to test the significance of
diffusion and of overshoot by themselves. In previous studies, models
that include helium diffusion during solar evolution were found to
improve the predicted p-mode frequencies for some modes and worsen the
agreement for others (Guenther \ea 1993). Here we verify this result
for both the Bahcall and Loeb (1990) formulation of diffusion and the
Proffitt and Michaud (1991) formulation of diffusion. We find that the
effects of rotation on the Sun's structure in the outer layers
perturbs the $p$-mode frequencies only slightly when compared to the
more substantial effects due to diffusion. In the thin overshoot layer
(taken here to be $0.1\, H_p$), we have compared the effect of
overmixing in a radiative layer versus convective (adiabatic)
penetration. Neither radiative overmixing nor adiabatic penetration
has any significant effect on the $p$-modes, probably in part because
the overshoot layer is constrained to be thin. The predicted neutrino
flux in our diffusion plus rotation model is 7.12 SNU for Cl
detectors, 127 SNU for Ga detectors and $5.00\times 10^6\,{\rm erg \,
cm^{-2} }$ for the $^8$B neutrinos; this is approximately half-way between
the standard solar model without diffusion, and the standard solar
model with diffusion alone.
\end{abstract}

\vspace*{20pt}
\begin{center}
{\large to appear in {\it The Astrophysical Journal\/} June 10, 1995}
\end{center}
\vspace*{-18cm}
\hspace*{14cm}
CITA--94--54

\section{Introduction}
Standard solar models now predict a $p$-mode
oscillation spectrum which agrees with the observed oscillation
spectrum of the Sun (Guenther \ea  1992a,b; Guzik \& Cox 1993),
within the estimated uncertainties of the physical input. This is the
result of many improvements in the input physics, most significantly
the advances in astrophysical opacities (Kurucz 1991; Rogers \&
Iglesias 1994). The importance of diffusion processes, not normally
included in the standard solar model calculation, have also been
explored in solar models (Proffitt \& Michaud 1991, hereafter PM;
Guenther \ea 1993; Guzik \& Cox 1993; Bahcall \& Pinsonneault 1992a
\& b). The diffusion of helium from the surface convection zone into
the radiative layers just below the convection zone was found to
modify the predicted frequencies of $p$-modes that are most sensitive
to the conditions near the base of the convection zone
(Christensen-Dalsgaard \ea  1993; Guenther \ea 1993; Guzik \& Cox
1993). Specifically helium diffusion was found to bring the
frequencies of $p$-modes with $\ell =$ 30 to 50 into closer agreement with
observations. Furthermore, the depth of the convection zone of the
solar models that include diffusion is closer to the depth derived
from inversion of the $p$-mode frequency observations
(Christensen-Dalsgaard \ea 1991) than standard solar models that do
not include diffusion.

The agreement is not perfect, though. Even better agreement was found
when the efficiency of diffusion (Bahcall \& Loeb, 1990) was reduced
by a factor of two. In addition, it appears that the improvement in
the $p$-modes with $\ell$-values from 30 to 50 is at the expense of other
$p$-modes. That is to say, including diffusion affects adversely the
agreement for modes that penetrate more deeply, $\ell =$ 0 to 30, and modes
that do not penetrate to the base of the convection zone at all,
$\ell =$ 50 to 100 (Guenther \ea  1993).

In parallel with our work on diffusion in the Sun, we have studied the
rotational history of the Sun following the approach of Endal \& Sofia
(1978, 1981). This work has been carried out using a version of the
Yale code that includes some of the effects of rotation, such as,
angular momentum transfer and the associated chemical mixing due to
rotational instabilities (Pinsonneault \ea  1989, 1990). In this
approach, the radial dependence of the angular velocity is specified
in convective regions; in the fully convective pre-main sequence
phase, this plus the total angular momentum gives the initial
conditions. Angular momentum is assumed to be lost from the surface
convection zone of the models due to a magnetic stellar wind. In
radiative regions, angular momentum is conserved locally; the
timescales for angular momentum transport and mixing are then
estimated and solved for, using a coupled set of diffusion
equations. In such models, the spindown of the outer convection zone
creates a shear layer at its base.

Rotational mixing will counteract element separation; this is
especially important below the convection zone where both mechanisms
are the most effective. In addition, diffusion will interact with
other mixing processes associated with the outward transfer of angular
momentum from the interior in the Sun. This non-linear interaction
between rotationally induced mixing and diffusion has recently been
modeled in a self-consistent way during solar evolution by Chaboyer
(1993) and Chaboyer \ea  (1994) (hereafter CDP). Earlier work by
Proffitt and Michaud had considered the competition between diffusion
and a hypothetical ``turbulent diffusion'' law meant to simulate
rotationally induced turbulence in the Sun.

The purpose of this paper is to further explore the effects of
diffusion on the $p$-mode frequencies. We wish to determine if the
effort to include a more realistic treatment of transport phenomena in
solar models leads to better predictions of the solar $p$-mode
oscillation frequencies. We also wish to investigate the importance
and the nature of the thin transition layer at the base of convection
zone. In particular, we are interested in testing whether this
transition overshoot layer is best represented by simple convective
overmixing (in radiative equilibrium), or by convective penetration
(in adiabatic equilibrium), as discussed by Zahn (1991, 1993).

Finally we estimate the effect of these improvements on the calculated
neutrino flux. Bahcall and Pinsonneault (1992a \& b) have computed
solar models with the Yale code to evaluate the expected flux of solar
neutrinos, and have explored the impact of helium diffusion and other
changes in the input physics for the solar model. In the case of the
neutrino flux, the discrepancy between models and observations (the
``Solar Neutrino Problem'') has not only remained, but has
increased (Turck-Chi\`{e}ze \ea  1988; Sackmann, Boothroyd and Fowler
1990). Proffitt (1994) including both helium and heavy element
diffusion in his calculations, and using a different set of diffusion
coefficients, has found an even larger increase in the expected
neutrino fluxes.

The solar models presented here were constructed with the rotating
version of the Yale stellar evolution code to which was added the
treatment of thermal diffusion in the Bahcall and Loeb (1990, BL) and
Proffitt and Michaud (1991) formulations. The constitutive physics
differs from the Bahcall-Pinsonneault (1992a \& b) models only in some
details, such as the treatment of the Debye-H\"{u}ckel corrections to
the equation of state (see Guenther \ea  1992). The models are fully
described in Chaboyer (1993) and CDP. In section 2, we briefly
describe the set of models calculated. In section 3 we compare the
observed $p$-mode spectrum of the Sun to the standard solar model and
the altered solar models that include a variety of combinations of the
following physical processes: Bahcall and Loeb diffusion, Proffitt and
Michaud diffusion, rotation, and the thin overshoot layer below the
convection zone. The neutrino fluxes are discussed in section 4.

\section{The Models}
The Sun is depleted in lithium by a factor of 100 to
200, and beryllium by 0.3 dex, with respect to the meteorites (Anders
\& Grevesse 1989). To help explain the observed lithium surface
abundances, it has been suggested that turbulent mixing takes place
during evolution below the convection zone (Schatzman 1969, 1977;
Vauclair \ea 1978; Vauclair 1988). Deliyannis (1990) has emphasized
the importance of explaining both lithium and beryllium depletion
simultaneously by the same mixing process. Such mixing is needed even
when the effects of diffusion of light elements are taken into
account. It has also been pointed out that the presence of a fully
mixed layer at the base of the convection zone, due to convective
overshoot or penetration, could be detected using solar seismology
(Berthomieu \ea  1993; Monteiro \ea  1993; Stix 1994). In rotating
solar models (Pinsonneault \ea  1989, 1990), the existence of a
mixed layer follows naturally from the large rotational shear at the
base of the convection zone which generates turbulence, and in turn
mixes the shear layer. In this work we examine the significance of
diffusion, rotation, and the mixed layer within the context of the
$p$-mode oscillation spectrum. To carry out this investigation we have
constructed a variety of models that include some combination of the
effects of rotation, diffusion, and a mixed layer. In addition,
because some authors model the overshoot layer by extending the
adiabatic temperature gradient below the base of the convection zone
we also calculate solar models with this formulation of overshoot.
Specifically we have constructed solar models with the following
characteristics:
\begin{enumerate}
\item ssm  --- a standard solar model,
\item  nodh --- a standard solar model without the Debye-H\"{u}ckel
correction to the equation of state,
\item  adov01 --- a standard solar model
with a thin ($0.1\,H_p$) mixed adiabatic layer below the convection zone,
\item adov03 --- a standard solar model with a thin ($0.3\,H_p$) mixed
adiabatic layer below the convection zone,
\item mx01 --- a standard solar model with a thin ($0.1\,H_p$) mixed
radiative layer below the convection zone,
\item  pm --- a standard solar model with helium diffusion in the
Proffitt and Michaud formulation,
\item  pm08 --- a standard solar model with helium diffusion in the
Proffitt and Michaud formulation but with the diffusion efficiency
reduced by 20\%,
\item  bl --- a standard solar model with helium diffusion in the
Bahcall and Loeb formulation,
\item d+r  --- a standard solar model with rotation
plus helium diffusion (Proffitt and Michaud formulation but with the
diffusion efficiency reduced by 20\%),
\item  d+rZ --- a standard solar
model with rotation plus helium diffusion as in model d+r, but with
the same Z as model pm08.
\end{enumerate}

The standard solar model and its variations are nearly identical to
the models described in CDP. Two modifications were made to
accommodate the $p$-mode frequency calculations: (1) a detailed
atmosphere based on the Krishna Swamy (1966) empirical fit to the
solar atmosphere was added to the models, and (2) the inner most mass
shell was positioned closer to the actual center (required for
accurate $\ell = 0\,p$-mode frequency calculations).

The models and some of their key distinguishing characteristics are
summarized in Table 1. The mixing length and helium abundance of
each model have been adjusted so that the radii and luminosities of
the models match each other to one part in $10^6$
($L_{\sun} = 3.8515\times 10^{33}\,{\rm erg\,s^{-1}}$;
$R_{\sun} = 6.9598\times 10^{10}\,{\rm cm}$ )
at the common age of 4.55 Gyr. The solar age from the zero-age main
sequence has been estimated by Guenther (1989, revised in 1992a) to be
$4.52\pm 0.03$ Gyr. All the models except the adov01 and adov03
models were evolved from the fully convective pre-main sequence
phase. They reached the main sequence in about 0.04 Gyr.

The models constructed with a mixed adiabatic layer extending below
the base of the convection zone (adov01 and adov03) were evolved from
a ZAMS model (produced during the ssm evolutionary run). This
introduces a small error of the order of one part in $10^5$ in the
temperature and density because of some loss of information in the
entropy term. The effect on the $p$-mode spectrum is negligible. But
it should be kept in mind that there is a small systematic difference
in the calculated neutrino flux of $+0.04$ $^{37}$Cl SNU compared to evolving
from the pre-main sequence. We did not evolve these models from the
pre-main sequence because the numerical algorithm to extend the
adiabatic temperature gradient below the surface convection zone lead
to convergence problems as the model ceased being fully
convective. This difficulty is only encountered when we impose the
high tolerances (few parts in $10^7$) necessary for solar $p$-mode
calculations, and is not present if we relax the tolerances to
standard stellar values. The question of how to properly treat the
ad-hoc inclusion of adiabatic overshoot during pre-main sequence
evolution will be investigated separately. Here we avoid the question
all together by starting our evolution for these two models on the
ZAMS.

\section{The $P$-Mode Frequencies}
\subsection{Introduction}
 Diffusion affects the
structure of the solar model in three ways: (1) it increases the
helium abundance in the interior and decreases it in the convective
envelope; (2) it increases the depth of the convective envelope
(Guenther 1994) and (3) it changes the surface value of Z since Z/X is
fixed by observation. Those $p$-modes whose inner turning points are
near the predicted base of the convective envelope, $\ell =$ 30 to 50, (see,
e.g., Fig.1 in Guenther 1994) will be the most sensitive to the
effects that diffusion has on the position of the base of the
convection zone. The frequencies of lower $\ell$-valued $p$-modes, which
penetrate deeper will be affected by the increased interior helium
abundance, and the higher $\ell$-valued $p$-modes, which probe only inside
the convection zone, will be affected by the decreased helium
abundance of the convective envelope.

The frequencies of $p$-modes with $\ell$-values less than 100 were
calculated for the set of models. In Figure 1 we show the frequency
difference between the models and the observations (Libbrecht \ea
1990). The mx01 (the standard solar model with an over mixing layer
extending $0.1\,H_p$ below the convection zone) is not shown because the
model is identical in all respects to the ssm model. The d+rZ model is
not shown as it was only constructed to test our understanding of the
effects of rotation and metallicity on the neutrino fluxes. Lines
connect $p$-modes with common $\ell$-values (not all $\ell$-values are plotted,
only a representative sample). Bold lines are drawn for $\ell =$ 30, 40, and
50; these modes are very sensitive to the conditions near the base of
the convection zone. If the $p$-mode spectrum of a model were
identical to the observed spectrum then the plot would appear as a
series of overlapping horizontal lines passing through
$\nu_{\rm model} - \nu_{\rm s} = 0\,\mu$Hz.

The agreement between theory and observation is not perfect. Errors in
the modeling of the superadiabatic layers and the surface boundary
conditions are responsible for the non-horizontal slope of the lines
(Guenther 1994). Errors in the surface layers affect the frequencies
of all modes, primarily by altering the average spacing between
adjacent in $n$ modes. This is seen as a non-zero slope in the frequency
difference diagram.  In the model this error is associated with the
difficulties of modeling turbulent convection, calculating low
temperature opacities that include the effects of molecules, and
modeling the surface boundary condition. For the purposes of this
work, we will ignore these problems and focus on errors that we can
actually address.

Even when we ignore the non-zero slope of the lines there still
remains the error associated with the scatter of the lines, i.e., the
thickness of the bundle of lines. Modes with different $\ell$-values
penetrate to different depths, hence, the fact that lines joining
different $\ell$-value modes do not lie on top of each other tells us that
the interior model is not correct to varying degrees at different
depths. Improvements to the modeling of the interior, via opacities,
equation of state, inclusion of diffusion, etc., should reduce the
spread or thickness of the bundle of iso-l $p$-mode frequency lines.

\subsection{Debye-H\"{u}ckel Correction and the Significance of the
Perturbations}
A casual comparison of the thickness of the bundles of
lines in the plots of FigureJ1 reveal little difference among the
models. The effects of diffusion, overshoot, and rotation are not that
large, producing shifts in the frequencies of at most $\pm 3\,\mu$Hz. The
biggest difference appears between Figure 1a and 1b. Figure 1a
corresponds to our standard solar model, i.e., our reference model for
this work. Figure 1b corresponds to a solar model identical to our
standard solar model in all respects except that the Debye-H\"{u}ckel
correction has not been applied to the equation of state
calculation. Improvements to the equation of state and the opacities
have already been shown (Guenther \ea  1992; Christensen-Dalsgaard,
D
1990;Christensen-Dalsgaard 1988; Ulrich \& Rhodes 1983) to lead to
improved $p$-mode frequency calculations. Here we use the model
lacking the Debye-H\"{u}ckel correction not only to emphasize its
importance in producing accurate solar models for $p$-mode frequency
determinations but also to provide a reference against which we will
judge the significance of the effects of diffusion, overshooting, and
rotation on the $p$-mode frequencies.

Clearly the effects of diffusion and rotation as indicated by the rest
of the plots in FigureJ1 are not as large as the Debye-H\"{u}ckel
improved equation of state (the nodh model is the only model that does
not include the Debye H\"{u}ckel correction). Therefore, all
conclusions that we draw regarding the beneficial or detrimental
aspects of diffusion, and rotation, as implied by the $p$-mode
oscillation spectrum, will not be as strongly supported as is the case
for the Debye-H\"{u}ckel correction.

\subsection{Overmixing and Convective Penetration}
We have noted the
importance of the existence of a thin mixed layer at the base of the
convection zone, which is needed in the rotation models to describe
the shear layer.  Since there is some question about the effective
mean temperature gradient in this layer, we have considered two
extreme cases. In the first case, we simply have allowed the mixed
layer to be chemically homogenized beyond the base of the convection
zone by $0.1\,H_p$, and have let the Schwarzschild criterion determine
whether the local temperature gradient is radiative or convective
(overshooting in the terminology of Zahn 1991). Since the opacity is
unaffected by the mixing in this case, the layer is radiative (model
mx01).  In the second case, we have extended the convection zone by
applying the convective (adiabatic) temperature gradient by $0.1\,H_p$
and $0.3\,H_p$ beyond the formal boundary of the convection zone
(models adov01 and adov03, respectively). See Figure 2. This
procedure is often called convective overshoot (Ahrens \ea  1992),
and is referred to as convective penetration by Zahn (1991,1993).

The model with an artificially mixed layer below the base of the
convection zone model mx01) is identical, within the numerical
accuracy of the model, to the standard solar model, and as a result
the $p$-mode frequencies are also identical to standard model $p$-
mode frequencies. This is expected because the region below the solar
convection zone is chemically identical (except for trace amounts of
Li and Be) to the convection zone. This is not the case in models that
include diffusion, where, as a consequence, overmixing can lessen the
penetration of He below the convection zone. We note that larger
overshoot (by as much as $0.3\,H_p$) is required to match the solar Li
abundance in non-rotating non- diffused solar models (Ahrens \ea
1992) by enhanced pre-main sequence burning. But overshoot above
$0.1\,H_p$ seems ruled on other grounds, such as, the observations of Li
abundances in open star clusters which contain young solar analogs
(Chaboyer \ea   1994b).

The $p$-mode frequencies of the solar model in which the adiabatic
temperature gradient is extended $0.1\,H_p$ below the base of the
convection zone (adov01) are nearly identical to the $p$-mode
frequencies of the standard solar model. Not until the adiabatic
overshoot reaches $0.3\,H_p$ do we see a significant affect on some of
the $p$-modes.  Comparing Figure 1d with 1a we see that only the
$\ell = 30$ to 50 $p$-modes are affected. The structure near the base
of the convection zone (see Figure 2) of the adov03 model is
sufficiently different from the ssm model to perturb the frequencies
of the $p$-modes that probe this region. We thus conclude that we are
unable at this time, within the present uncertainties, to
differentiate between radiative overmixing and convective penetration
of the order of $0.1\,H_p$ in our models on the basis of the $p$-mode
spectrum. Larger amounts of adiabatic overshoot (i.e., $\ge 0.3\,H_p$)
do reveal themselves in the $p$-mode spectrum but are ruled out by the
constraints set by the observations of lithium destruction in the
early phases of the evolution of the Sun. Our conclusion differs from
that reached by Berthomieu \ea  (1993), who found an improvement in
the $p$-mode spectrum of their convective penetration model over the
standard solar model. Comparing their ``delta frequency'' plots
with our plots (Figure 1) reveals that the $p$-modes of their
standard reference model are a factor of two more discrepant with
observation than our standard reference model (ssm). We speculate that
the substantial overshoot layer which they introduced may have
compensated for the shortcomings of their reference solar model.  This
example illustrates the pitfalls in interpreting differences between
theoretical models and observation when many unknown sources of error
in the model remain.

\subsection{Diffusion}
Comparing Figures 1e (model pm), 1f (model pm08) and 1g
(model bl) with Figure 1a (model ssm), it does not appear that
diffusion improves the solar model at all.  In fact, the spread of
iso-$\ell$ lines is greater. Looking only at the $\ell =$ 30, 40, and 50 lines
(heavy stroked lines) we see that the scatter of these lines is
reduced. Because the inner turning points of these modes are near the
base of the convection zone, they are maximally sensitive to the
position of the base of the convection zone --- modes turning below
the base are maximally perturbed by the abrupt change in the
temperature gradient while those turning above are not. When the
frequencies of these modes are compared to the frequencies of the Sun,
an errant location in the base of the convection zone in the model is
immediately discerned by the markedly greater spread of the $p$-mode
frequency differences. For a model with a poorly positioned convection
zone base, the iso-$\ell$ lines for the $\ell = 30$ to 50 $p$-modes will be
tightly bundled at low frequencies where all the $p$-modes have
turning points above the base of the convection zone and then will
flay apart ($\ell = 30$ first, then $\ell = 31$, etc.) as the frequency
increases and the turning points begin to dip below the base of the
convection zone.

The location of the base of the convection zone in modes that include
diffusion of helium, as noted by Guenther \ea  (1993) and Guzik \&
Cox (1993), more closely agrees with the depth implied by inversion
(Christensen-Dalsgaard \ea  1991), hence, so do the $\ell = 30$ to 50
$p$-modes. Unfortunately, not all the modes are improved. In fact, the
overall scatter of the lines, i.e., the bundle thickness for $\ell =0$ to
100) is increased. There is still something wrong with this solar
model. Guzik and Cox (1993) have succeeded in improving the agreement
between model and observation beyond that represented here, by
adjusting the opacities in their models. We believe that errors in the
opacities just below the convection zone and in the outer layers
and/or errors in the solar chemical abundances (within the present
quoted uncertainties), may be responsible for most of the remaining
discrepancy.

The two different formulations of diffusion tested here are
insignificantly different, with regard to the $p$-mode frequencies. If
Figures 1e, 1f, and 1g are superimposed on top of each other then
slight differences can be seen, but the magnitude of the differences
are close to the magnitude of the numerical error associated with the
model and pulsation calculation, which is approximately $\pm 0.3\,\mu$Hz
(Guenther \ea  1989).  The pm08 model is identical to the pm model
except the diffusion coefficients of the pm08 are scaled to 80\% those
in the pm model. We note that the frequencies of the $\ell =30$ to 50
$p$-modes of the pm08 model are nearly identical to the bl model and
that the frequencies of the $\ell =30$ to 50 $p$-modes of the pm08 and
bl models are more tightly bundled than the pm model. Because of the
numerical uncertainties in the model calculation we do not attribute
any significance to this.

\subsection{Rotation Plus Diffusion}
The $p$-mode frequencies of the models
that include rotation and diffusion (see Figure 1h; model d+r) are
insignificantly different from the models that include just diffusion
(compare Figures 1e, 1f, and 1g). When compared to the standard
solar model (Figure 1a) the overall effect on the modes, like that
of diffusion by itself, is to worsen the agreement. Despite the
tremendous amount of effort required to produce the rotation plus
diffusion models, we must conclude that rotation is not the missing
ingredient that when included in the solar model calculation removes
the remaining discrepancies between model and observation. Rotation
hardly affects the internal structure and the $p$- mode frequencies at
all.

However, rotation does tend to inhibit the diffusion, i.e., a model
with rotation and diffusion is similar to a pure diffusion model with
approximately one half the efficiency of diffusion (CDP). By
intimately connecting the effects of rotation with diffusion (CPD) one
perturbs the effects of diffusion on the solar model structure. The
efficiency of diffusion is effectively reduced by the shear induced
turbulence below the convection zone. Without rotation and the
associated shear layer, helium diffuses out the base of the convection
zone and remains within a diffusion scale length of the base
($~ 0.05\,R_{\sun}$). With rotation, the shear induced turbulence below the
convection zone spreads out the helium that has diffused out of the
convection zone into deeper layers and it pushes some of the helium
back into the convection zone, thereby reducing the effective
efficiency of diffusion. Contrary to previous conclusions
(Christensen-Dalsgaard \ea 1993; Guzik \& Cox 1993), the presence of
a (possibly rotationally induced) mixed layer due to turbulent
diffusion just below the convection zone does not necessarily affect
adversely the calculated $p$-mode spectrum. The effect on the
$p$-modes depends only slightly on the adopted formulation of the
diffusion coefficient.

\section{Rotation Curve}
While the approach of using the $p$-mode
frequencies described in preceding sections provides a sensitive test
of the effects of internal rotation on solar structure, it says little
about the rotation curve itself. This is primarily because rotation is
slow in these models and barely modifies their internal structure. But
the distribution of angular momentum in the present Sun can be tested
directly by observations of rotational splittings of oscillation modes
which penetrate to different depths in the Sun, and with different
amplitudes. Recent inversions of the lowest l modes observed by the
IPHIR space mission (Toutain \& Fr\"{o}hlich 1992), and of the IRIS
network data (Loudagh \ea   1993) yield a rotation rate at $0.2\,R_{\sun}$
which is higher than the surface value and might be as high as 4 times
the observed solar surface value, which is compatible with some of the
models constructed by CPD. On the other hand, the inversion of
intermediate $\ell$ $p$-mode splitting data (Libbrecht \ea  1990) yields
a flat (solid body) rotation curve from the surface to $r \approx 0.4\,
R_{\sun}$. Even our minimal differential rotation model (see rotation curve
in Fig. 3 and discussion below) predicts a rotation rate in the
radiative envelope which is too high by at least a factor of two when
compared to the rotation rate inferred $p$-mode splittings in the
Sun. This conclusion is corroborated by a recent measurement of the
solar oblateness (Sofia \ea  1994) which puts an independent limit
on the amount of angular momentum in the outer parts of the solar
radiative envelope which is 2 sigma below the oblateness of the
minimal difrerential rotation model.

Another approach which has proved fruitful in testing models of the
evolution of rotating stars is to appeal to observations of solar
analogs at different stages of evolution.  Observations of the surface
rotation periods of solar analogs more evolved than the Sun and now
observed as subgiant stars, are best explained if angular momentum
buried deep in their interiors is being dredged-up into their
convective envelopes as their convection zones deepen (Pinsonneault \ea
 1989; Demarque \& Guenther 1988). There is in addition evidence
from surface rotation velocities, that evolved Sun-like stars have
preserved a significant fraction of their initial angular momentum in
their interiors into advanced phases of evolution (Pinsonneault \ea
1991). If these objects are true solar analogs, we must expect
that there is similarly a reservoir of angular momentum in the deep
interior of the Sun which is now buried in the inner 50\% of its radius
(or 90\% of its mass). Some of this angular momentum will eventually be
dredged up to the surface from the deep interior as the Sun evolves
into a subgiant. Since all other evidence indicates that the outer
radiative layers of the present Sun rotate slowly, this angular
momentum must be located within the inner 50\% of the solar radius. The
rotation profiles of the models shown in Figure 3 exhibit too much
rotation between 50\% and 70\% of the radius.

The minimal differential rotation model shown in Fig. 3 was
calculated assuming that the inhibitive effects of the mean molecular
weight gradient can essentially be ignored and that the GSF rotational
diffusion coefficients (James \& Kahn 1970,1971) are ten times larger
than the default values used in the d+r model. The default values give
the best fit to the observed rotation velocities, and Li abundances of
young cluster stars. This minimal differential rotation model, which
represents an extreme case, indicates that some important physical
process or processes, which are responsible for removing some of the
angular momentum from the outer radiative layers of the Sun, are
missing in our formulation.

\section{Predicted Neutrino Flux}
The structural effects of rotation are
small, and so rotation by itself does not appreciably alter the
predicted neutrino fluxes in solar models (cf. Bahcall 1989).
However, recent studies have found that diffusion of $^4$He increases the
predicted solar neutrino fluxes by approximately 11\% for the Cl and
water detectors, and 4\% for the Ga detectors (Bahcall \& Pinsonneault
1992a \& b; Proffitt 1994). Rotational induced mixing counteracts the
effects of diffusion, and so models which include the combined effects
of rotation and diffusion predict neutrino fluxes between standard
models (no diffusion), and pure diffusion models. This is shown in
Table 1. We note that the neutrino flux of our ssm model is
approximately 0.4 SNU for $^{37}$Cl lower than the best standard solar
model of Bahcall \& Pinsonneault (1992a). Our models use the nuclear
energy generation of Bahcall (Bahcall \& Pinsonneault 1992) and the
neutrino cross sections are also the same as those used in Bahcall \&
Pinsonneault (1992a). The difference in neutrino fluxes in the
standard models is primarily due to the different final ages (4.55 Gyr
versus 4.6 Gyr of Bahcall and Pinsonneault), which results in a 0.2
SNU decrease, and a combination of the different final luminosities
($3.8515\times 10^{33}\,{\rm erg \,s^{-1}}$ versus
$3.90\times 10^{33}\,{\rm erg \,s^{-1}}$ of Bahcall and
Pinsonneault) and different helium and heavy metal abundances (0.2708
and 0.01880 versus 0.2716 and 0.01895 of Bahcall and
Pinsonneault). There are also differences in the treatment of the
Debye-H\"{u}ckel correction (which here includes the effects of
electron degeneracy), the opacity interpolation procedure, and in the
chosen values of some of the physical constants.

The difference in solar neutrino fluxes predicted by our standard
models and pure diffusion models, $~0.8$ SNU increase for $^{37}$Cl, are in
good agreement with those found by Bahcall \& Pinsonneault (1992a \&
b) and Proffitt (1994). The combined models, which include rotation
and diffusion, predict solar neutrino fluxes which are between the
standard models and the pure diffusion models, with rotation reducing
the effect of diffusion on the neutrino fluxes by one-third. The last
row in Table 1 (model d+rZ) summarizes the characteristics of a
combined model in which Z has been adjusted to be the same as in the
pure diffusion model pm08, with $Z=0.01941$. We see that $0.18$ SNU (or
approximately half) of the reduction in the neutrino flux in d+r is
due to the change in $Z$. The rest of the reduction ($0.25$ SNU) comes
from the fact that the diffusion in the core is also being slowed
down.

Including the effects of heavy-element diffusion does not appreciable
effect the predicted $p$-mode frequencies, but has a comparable effect
to $^4$He diffusion on the solar neutrino fluxes (Proffitt
1994). Proffitt (1994) determined neutrino fluxes of 9 SNU for Cl
detectors, 137 SNU for Ga detectors and
$6.48\times 10^6\,{\rm cm^2\, s^{-1}}$ for the $^8$B
flux for models which include $^4$He and heavy-element diffusion, and
suggested that turbulent mixing could reduce these amounts by a
maximum of 0.17 SNU (Cl), 0.88 SNU (Ga) and
$0.13\times 10^6 \,{\rm cm^2\, s^{-1}}$
($^8$B). Inter-comparing no diffusion models, the pure diffusion model
and the combined rotation plus diffusion model in Table 1 indicates
that turbulent mixing due to rotation has a larger effect than that
suggested by Proffitt (1994). The turbulent mixing induced by rotation
will lead to reductions of 0.34 SNU (Cl), 1.8 SNU (Ga) and
$0.26\times 10^6 \,{\rm cm^2\, s^{-1}}$
in the predicted solar neutrino fluxes of Proffitt
(1994).

\section{Summary}
\subsection{Solar internal structure and dynamics}
  We have
considered the helioseismic properties of solar models which take into
account rotationally induced mixing and its interaction with
diffusion. We emphasize that rotating-diffusive models are very
similar in structure to standard models of the Sun. But they differ
from the standard models in several subtle but important ways (in
particular the depth of the convection zone and the structure and
composition profile of the radiative layers just below the convection
zone) which affects the $p$-mode frequencies.

We find that models that include diffusion (either the Bahcall-Loeb
formulation or the Michaud-Proffitt formulation) and models that
include diffusion and rotation predict $p$-mode frequencies that are
in better agreement with observation for only those modes sensitive to
the location of the base of the convection zone. Higher l $p$-modes,
confined to layers above the convection zone, are adversely
affected. The effects on the $p$-mode frequencies of rotation cannot
be distinguished from the effects of diffusion, although the latter
produces a larger effect. With regard to rotation, the $p$-modes are
only seeing the inhibitive effect that rotation has on diffusion via
the shear induced turbulent layer, and the diffusion coefficients can
be adjusted within their error to mimic the effect of rotation on the
$p$-mode frequencies. At the present time, the errors associated with
the calculation of the diffusion coefficients are too large to enable
us to say one way or the other whether rotation is a necessary
ingredient in the solar model. The diffusion coefficients can be
adjusted within their error to mimic the effects of rotation on the
$p$-mode frequencies.

On the other hand, as stellar astronomers we cannot ignore the
considerable success of the rotating models in interpreting
observations of other Sun-like stars of different ages (Pinsonneault
\ea 1989, 1990; Chaboyer \ea 1994b). These results argue
convincingly for the existence of a well-mixed transition layer in the
Sun, most plausibly the result of rotational shear. From the $p$-mode
frequencies, we are unable to distinguish between an adiabatic mixed
layer (convective penetration) as favored by Zahn (1991,1993), and a
radiative mixed layer for overshoot layers less than $0.1\,H_p$.

It is still very difficult to determine the rotation profile deeper
inside the Sun.  Here the observational errors are large, and the
models more uncertain. There is however agreement over the main result
that the inner part of the Sun rotates faster than its surface.  This
conclusion is in agreement with the preliminary inversions from the
IPHIR space mission (Toutain \& Fr\"{o}hlich 1992) and the IRIS network of
whole-Sun observing stations (Loudagh \ea 1993). All observational
evidence also points to slow rotation in the radiative layers just
below the convection zone, which indicates that there is a missing
angular momentum transfer mechanism in our models, which is very
efficient in flattening out the solar rotation curve, at least in the
outer part of the radiative envelope (see the rotation profiles in
Figure 3).

Even though the $p$-modes of our models do argue favorably for
diffusion plus overshoot, the remaining errors in the model (and their
affect on the $p$-mode frequencies) are large enough that if they
cannot be accounted for as errors in the opacity and equation of
state, then any combination of diffusion, overshoot, and overmixing,
which is consistent with the observed surface abundances, is
possible. We do not believe, though, this to be the case, and the
simplest interpretation lies in opacity errors. We know that the
opacities are accurate to only 115\% and that smaller adjustments to
the opacities can correct the $p$-mode spectrum of the solar
model. Guzik and Cox (1993) have already shown that by tweaking the
opacities, the $p$-mode frequencies of a solar model with diffusion
can be made to agree with the observations to almost $\pm 1\,\mu$Hz.

\subsection{Solar neutrinos}
 Because the rotating-diffusive models only differ
in small and subtle ways from the standard solar model, the predicted
solar neutrino flux is not greatly affected by the presence of
rotation and/or diffusion in the Sun. The results are summarized in
Table 1.  We see that the predicted neutrino flux is 6\% lower in the
rotation plus diffusion model compared to the diffusion only models.

Rotation by itself does not alter the predicted solar neutrino fluxes,
as the structural effects of rotation are small. However, the mixing
induced by rotation counteracts the effects of diffusion on the solar
neutrino fluxes, such that our rotation- diffusion models have
neutrino fluxes intermediate between the Bahcall \& Pinsonneault
(1992a \& b) standard and pure $^4$He diffusion models.

\acknowledgements
We acknowledge support for this research from grants NASA NAGW-2531,
NAG5-1486 and NAGW-2469 to Yale University. DBG acknowledges the
support from the NSERC. DBG thanks undergraduate student A. Forgeron
for bringing to our attention an error in one of our earlier
calculations of the overshoot models.

\appendix

\clearpage

\begin{centering}
\Large \bf FIGURE CAPTIONS \\
\end{centering}

\begin{description}
\item[FIG. 1:]
The frequency differences, model minus observed (Libbrecht \ea 1990)
of a sample of $p$-modes are plotted opposite the observed frequency
of the mode. Lines join common $\ell$-value modes for $\ell =$ 0, 1, 2,
3, 4, 10, 20, 30, 40, 50, 60, 70, 80, 90, and 100. The $\ell =$ 30, 40,
and 50 modes are connected by a thick line. Individual plots
correspond to the different solar models as identified in the text and
Table1.

\item[FIG. 2:]The temperature gradient as a function of radius
fraction is shown for the ssm, adov01, and adov03 models.

\item[FIG. 3:]The angular rotation velocity is plotted as a function
of interior mass fraction $M_r/M_{\sun}$, Fig 3(a), and radius
fraction $r/R_{\sun}$, Fig 3(b), for the specific rotating model used
in this paper (d+r or model VN of Chaboyer 1993), solid line, and for
a minimal differential rotation model (model SF of Chaboyer 1993),
dashed line.
\end{description}
\end{document}